\documentclass[twocolumn,showpacs,preprintnumbers]{revtex4}
\usepackage{amssymb}
\usepackage{amsmath}
\usepackage{graphicx}
\usepackage{epsfig}

\begin{document}

\title{\bf Point-contact spectroscopy of the phononic mechanism of superconductivity in YB$_6$}
\author{P Szab\'o,$^1$ J Girovsk\'y,$^1$ Z. Pribulov\' a,$^1$ J Ka\v cmar\v c\' ik,$^1$ T. Mori,$^2$ and P. Samuely$^{1}$}

\address{$^1$Centre of  Low Temperature Physics @ Institute of Experimental Physics, Slovak
Academy of Sciences, Watsonova
47, SK-04001 Ko\v sice, Slovakia\\
$^2$ Advanced Materials Laboratory, National Institute for Materials Science, Tsukuba, Ibaraki 305-0044, Japan}

\begin{abstract}

Lortz \textit{et al.} [Phys. Rev. B \textbf{73}, 024512 (2006)] have utilized specific heat and resistivity measurements as  "thermal 
spectroscopies"  to deconvolve the spectrum of the electron-phonon interaction in YB$_6$ assuming a major role of the low frequency phonon mode 
in mediating superconductivity. 
Here, we present direct point-contact spectroscopy studies of the superconducting interaction in this system. As a result 
the normalized superconducting gap reveals a strong coupling with $2\Delta/k_BT_c = 4$ and moreover the spectra contain nonlinearities typical 
of the electron-phonon interaction at energies around 8 meV.   
The in-magnetic-field measurements evidence that the phonon features found in the second derivative of the current-voltage characteristics are 
due to the energy dependence of the superconducting energy gap as their energy position shrinks  equally as the gap is closing. This is a direct 
proof that the superconducting coupling 
in the system is due to the low energy Einstein-like phonon mode associated with the yttrium ion vibrations in a perfect agreement 
with determinations from bulk measurements. 

\end{abstract}

\maketitle

\section{Introduction}

Surprising discovery of superconductivity in MgB$_2$ at almost 40 K \cite{akimitsu} has re-attracted attention to the superconductors with an 
electron-phonon interaction mechanism. 
Due to their eight phonon branches and a high density of scatterers in the form of boron atoms the metallic and non-magnetic hexaborides were regarded 
once as possible candidates for high-temperature superconductors \cite{arko}. These expectations have not been met when YB$_6$ has the highest transition temperature \textit{T}$_c=$ 8.4 K 
among them \cite{buzea} and isostructural LaB$_6$ with a very similar  electronic structure is not superconducting at all \cite{batko}. 
It was believed for a long time that a dominant contribution to the electron-phonon interaction (EPI) in hexaborides leading eventually to superconductivity came
 from the boron sublattice with a lot of phonon branches stretched up to 160 meV \cite{schell}.  But later, Mandrus \textit{et al.} \cite{mandrus} noticed 
that due to a large space of metal atom among the boron octahedral cages the metal atoms can develop large unharmonic vibrational amplitudes with strong EPI. Then, LaB$_6$ 
or YB$_6$ can be modeled as a Debye solid of the rigid boron framework and metals ions can be treated as independent harmonic oscillators (Einstein 
oscillators). The local vibrational modes of the La ions are the most important for explanation of the low temperature resisitivity and heat capacity in 
LaB$_6$ as was also suggested by the neutron experiments of Smith \textit{et al.} \cite{smith} showing rather rapid flattening of the acoustic modes near 
13 meV due to non-interacting vibrations of La ions. 
Some of us \cite{samuely} have measured the EPI function of LaB$_6$ directly by the point-contact spectroscopy (PCS) detecting the whole set of 8 modes starting from the prominent peak at 13-15 meV up to 
160 meV with the resulting electron-phonon coupling constant  $\lambda \approx 0.15$ showing very weak electron-phonon coupling in the system. 
In case of YB$_6$  the rattling motion of Y ion is now supposed to be dominant in the electron-phonon coupling. Based on this assumption 
Lortz \textit{et al.} \cite{lortz} have exploited the specific heat and resistivity measurements as  "thermal  spectroscopies" to deconvolve the phonon
 density of states and the spectrum of the electron-phonon interaction in YB$_6$.  Their results suggest that the superconductivity is mainly driven by a 
low lying phonon mode (at $\approx 8$  meV) which is associated with the yttrium ions in oversized boron cages.
Lower vibration frequency of lighter Y in YB$_6$ than frequency of  La vibrations in LaB$_6$ has been attributed to the weaker bond  of Y due to smaller radius in the same boron cage.
The electronic specific heat revealed that
 YB$_6$ is a strong coupling  superconductor with the reduced energy gap $2\Delta/k_BT_c \approx 4.1$ and the coupling constant $\lambda \approx 1$. 
 Three spectroscopy experiments  performed on YB$_6$ have been published
 to our knowledge so far. Kunii \textit{et al.} \cite{kunii} determined from GaAs point contacts on YB$_6$ the superconducting gap $\Delta = 1.22$ meV 
which with the bulk $T_c = 7.1$ K (local $T_c$ was not measured) yields $2\Delta/k_BT_c \approx 3.8$. Also an EPI peak at 11 meV was observed with the gap 
energy subtracted. The photo-emission spectroscopy results of Souma et al. \cite{souma} indicate similar conclusions. Schneider \textit{et al.} 
\cite{schneider} prepared the tunneling sandwich on YB$_6$ film naturally oxidized with In top electrode and obtained $2\Delta/k_BT_c \approx 4$, 
the intense EPI peak at 8 meV in discrepancy with Kunii and $\lambda \approx 0.9$. No significant contribution to EPI was detected above 16 meV.

We present a detailed experimental study on YB$_6$ single crystal with $T_c$ of 7.5 K via point-contact (PC) spectroscopy.  Single 
$s$-wave superconducting energy gap with a reduced value of $2\Delta/k_BT_c$ close to 4 together with
 its classical temperature and magnetic-field dependence have been found. Moreover, an electron-phonon-interaction peak has been directly observed
in the second derivative of the current-voltage characteristics in the superconducting state. Upon application of magnetic field the energy position of this 
peak shifts to lower energies.
From data analysis the magnetic field dependence of the superconducting energy gap has been inferred. Importantly, the energy position of the 
EPI feature shifts in increasing magnetic 
field to lower energy exactly likewise the  superconducting gap is closing. 
This has been a direct proof that the low energy phonon mode near 7.6 meV is mediating superconducting pairing, in agreement with the 
conclusions of Lortz \textit{et al.}

\section{Point-contact spectroscopy of strong coupling superconductors}

A micro-constriction between two metals with the contact diameter $d$ much smaller than the mean free 
path of electrons $l$ can serve as a device for quaisparticles' spectroscopy since applied voltage $V$ is directly related to the 
quasiparticle energy excess $\Delta E=eV$, where $e$ is the electron's charge. 

In case of two normal metals forming the junction
the  PC current $I$ comprises beside the major term $V/R_{N}$, where $R_{N}$ is the PC
 resistance  also a small negative inelastic contribution $\delta I^{N}_{ph} (V)$ on the order of $\cong d/l_{in}$,
 where $l_{in}$ is the electronic inelastic mean free path, yielding nonlinearities in $I-V$ curve at characteristic EPI energies/voltages.  Small nonlinearities are better 
pronounced as peaks in the second derivative  $d^2V/dI^2(V)$ which is directly related to the  point-contact form of the electron-phonon interaction function
$g_{PC} = \alpha_{PC}^2(\omega)F(\omega)$ \cite{kos,naidyuk}. Here,  the matrix element $\alpha_{PC}(\omega)$ describes the strength of 
electron-phonon interaction in the PC geometry and $F(\omega)$ is the phonon density of states. 

When one of the PC-forming electrode is a superconductor,  below $T_c$ a phase coherent state of Cooper pairs is formed
in it.  For a bias energy $|eV|  < \Delta$, a direct transfer of
the quasi-particles is not possible due to existence of the energy gap $\Delta$  in the  spectrum of the superconductor.
 The transport of the charge carriers is realized through Andreev reflection with an excess current $I_{exc} (V) $ which 
makes the PC current inside the gap voltage twice bigger than in the normal state.  For biases larger than the gap voltage 
the excess current becomes constant and equals to
$I_{exc} \propto \Delta/R_N $. 
The PC conductance $\sigma =  dI/dV$ shows a double increase below the gap voltage $|V| < \Delta /e$ compared to the normal state or to what is
 observed at very large bias where
the coupling via the gap is inefficient. If a barrier is formed at the point-contact junction a Giaever-like tunneling component contributes to 
the charge transfer as well. Evolution  of 
the $dI/dV$  vs. $V$ curves  for different
interfaces  characterized by arbitrary transmission probability $T$ has been modeled
by Blonder, Klapwijk and Tinkham (BTK) theory \cite{btk}.
 In  case of a PC interface with an intermediate transmission probability $0<T<1$ a minimum appears  at zero bias  $eV$ = 0, but also two  peaks
 are visible at $eV \sim \pm \Delta/e$.   The experimentally measured PC conductance  data  can  be compared with  this model using  as input 
parameters the
energy  gap $\Delta$,  the  parameter  $Z$ (measure  of  the interface barrier strength  with  transmission
coefficient  $T  =  1/(1+Z^2)$), and a parameter   $\Gamma$  for   the  quasi-particle   lifetime broadening of the spectrum \cite{plecenik}.
The fitting procedure is described for example in Ref.\cite{szabo97}.

When a point-contact micro-constriction is formed between a normal metal and a strongly coupled superconductor, due to significant energy 
dependence of the superconducting gap $\Delta (eV)$ a small negative correction to the elastic excess current $\delta I_{exc} (V)$ caused by the Andreev reflection will
 make a measurable effect \cite{naidyuk}.  Generally, the PC current will read as

\begin{equation}
I(V) = \frac{V}{R_{N}} + \delta I^{N}_{ph} (V) + I_{exc} (V) + \delta I_{exc} (V), 
\end{equation}

where the first three terms are described above and the last term is given as  
$\delta I_{exc} (V) \cong (\Delta (V)/h\omega)^2$ \cite{om}, where $\omega$ is a characteristic frequency of phonons mediating the superconducting pairing. 
This term exceeds the inelastic component of the PC current and the electron-phonon interaction modes 
mediating superconductivity can be visible in the second derivative $d^2 V/dI^2 (V)$ as peaks not exactly at characteristic EPI energies, but, importantly, 
shifted to higher energies by the value of the superconducting energy gap $\Delta$, in the same way as in the tunneling spectroscopy. When the gap is closed 
for example by increasing temperature or in applied magnetic field, the EPI peaks should move to lower energies following the gap and decrease their intensity. This would be a smoking gun for the coupling mode 
which mediates the superconductivity in the system.

\section{Experiment}

The measurements presented in this paper were performed on high-quality single-crystalline YB$_6$  samples prepared by the traveling solvent floating 
zone method  \cite{otani}. All measurements were performed on crystals from the same batch. The crystals had a cubic form 
with the edge of about 0.5 mm. The values of the critical temperature $T_c$ have been determined by the point-contact-spectroscopy, resistivity and 
ac-calorimetry specific-heat measurements \cite{kacmarcik}. The first two techniques give $T_c = 7.4 \div 7.5$ K and the specific heat measurements sensitive to the bulk 
of the sample  yield  from the entropy balance construction around the anomaly $T_c$ = 7.32 K, a reasonably close value.  

The point-contact micro-constrictions were formed in-situ by pressing a Pt tip on the YB$_6$ surface using a differential
  screw  mechanism allowing for  a positioning of the tip on different spots on the sample. The PC tips have been cut off from 50 $\mu$m Pt wires. 
	Spectra with the best resolution have been obtained on the  shiny blue YB$_6$ surfaces prepared by cleaving
 before cooling down in the cryostat. PC spectroscopy measurements (the first and second derivatives of $I-V$ characteristics) have been 
realized by the standard lock-in modulation technique \cite{samuely}.  

\section{Results and discussion}

 PC resistances were typically in the range $R_N \cong 3 \Omega \div 30 \Omega$. 
As shown in our previous short communication \cite{giro1} in some point contacts with largest PC resistances a small gap below the BCS limit was observed. In those junctions also 
a non BCS temperature dependence of the gap had been obtained typical of the proximity effect from the bulk to the degraded surface layer. 
Here, we have collected a  large number of point contacts showing no signature of proximity effect and with a very little scattered gap size. 
The representative examples of the point-contact spectra $dI/dV$ obtained at $T = 4.2$ K are shown in Fig. 1. 
The lines display the experimental data after normalization to the normal-state PC conductance, measured above the transition temperature $T_c$ or 
above the upper-critical-field value $H_{c2}$.
The open symbols represent the fits by the BTK model. Although our experiments have been performed on freshly cleaved YB$_6$
 surfaces, the finite $Z$ parameter has always been found. Its size  
scattered as $Z \approx 0.3 - 0.8$. It means that there is always an effective interface barrier between the Pt tip and the sample. In some cases we 
were able to form point contacts with a high spectral resolution, as witnessed by a small value of the smearing parameter $\Gamma $ obtained
 from the fits. For the three spectra shown in Fig. 1 from top downwards the values of $\Gamma$ achieved 5, 13 and 37 $\% $ of the gap values,
 respectively.  The values of the superconducting gap obtained at 4.2 K on many junctions have been scattered between
 1.18 to 1.22 meV.

\begin{figure}[t]
\includegraphics[width=10 cm]{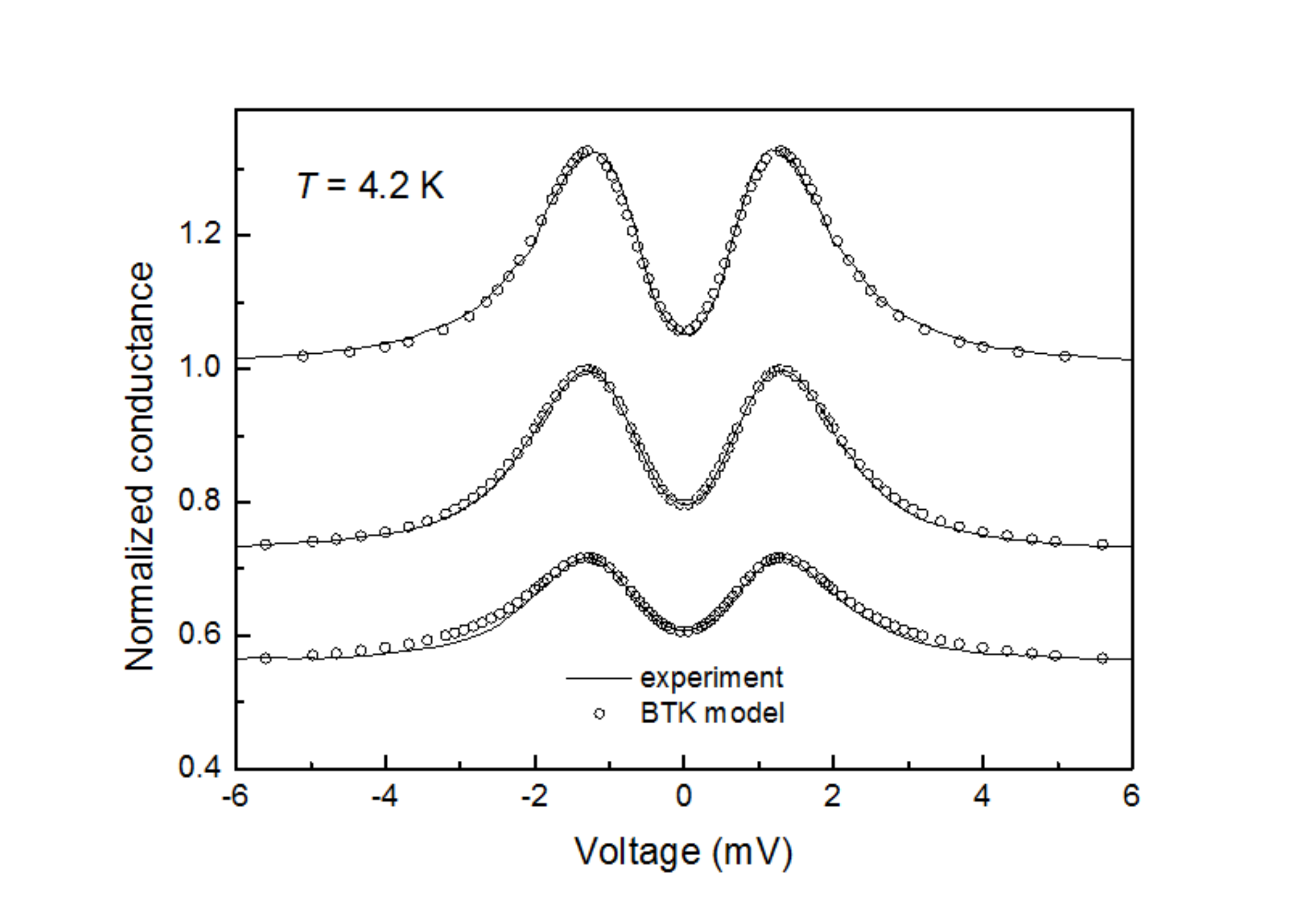}
\caption{Representative  point-contact spectra measured on Pt-YB$_6$ hetero-contacts at $T = $ 4.2 K  (solid lines). The lower curves are shifted in
 Y-coordinates for the clarity. Fits by the BTK conductances are shown by open symbols with the resulting parameters $Z =  0.65, 0.63, 0.64, \Gamma = 0.06, 0.16,
  0.45 $ meV  and $\Delta$(4.2 K) = 1.18, 1.22, 1.20 meV from top to bottom, respectively.}
\label{fig:fig1}
\end{figure}

Point-contact spectroscopy explores superconductivity in the area with dimensions on the order of the superconducting coherence length. In some cases
 the superconducting transition temperature $T_c$ at the surface can differ from the bulk value. That is why it is important to determine the local $T_c$
 of the junction before making any conclusions on such important parameter as the superconducting coupling strength $2 \Delta /k_B T_c$. In the published 
spectroscopy measurements on YB$_6$ the local critical temperature $T_c$ values have not always been determined, which could affect the calculation of the
 coupling strength. A correct determination of this ratio requires experimental measurement of $T_c$ and $\Delta $ in the same experiment. 

A representative temperature dependence of the PC spectra is plotted in Fig. 2 (lines). This is one of the high resolution spectra without any 
smearing parameter ($\Gamma = 0$).
The open circles are obtained from the fits to the BTK model. 
During the fitting procedure we first determined the parameters $\Delta $ and $Z$ at the lowest temperature, here equal to 1.6 K. Later, for 
higher temperatures only $\Delta$ was used as a fit parameter, while $Z$ we  had been kept constant. 
The inset of Fig. 2 shows the temperature dependence of the superconducting energy gap determined from the fit. The energy gap $\Delta (0) = 1.3$ meV 
closes at the critical temperature $T_c = 7.4$ K leading to the coupling strength of  $2\Delta/k_BT_c = 4.07 $.  Note that from a number of measurements
 we always obtained $T_c$  close to $7.4 \div 7.5$ K. The transition temperature  was determined either by noting when the PC conductance no longer displayed 
energy-gap features or  by extrapolating the temperature dependence of the gap to its zero value. The zero-temperature energy gap value was obtained 
either by extrapolating the temperature dependence of $\Delta$ to the zero temperature or by the measurements below 2 K where no extrapolation was
 necessary. As a result of many measurements we have obtained the superconducting gap $\Delta$(0) = 1.30 $\pm 0.03$ meV and $2\Delta/k_BT_c$ values between 3.9 and  4.1.
 We can conclude that in YB$_6$ the point-contact spectroscopy has revealed a single $s$-wave-gap superconductivity with intermediate strength of 
coupling. The temperature dependence of the superconducting gap follows  the BCS prediction well as documented by the full line in the inset of Fig. 2. 

\begin{figure}[t]
\includegraphics[width=10 cm]{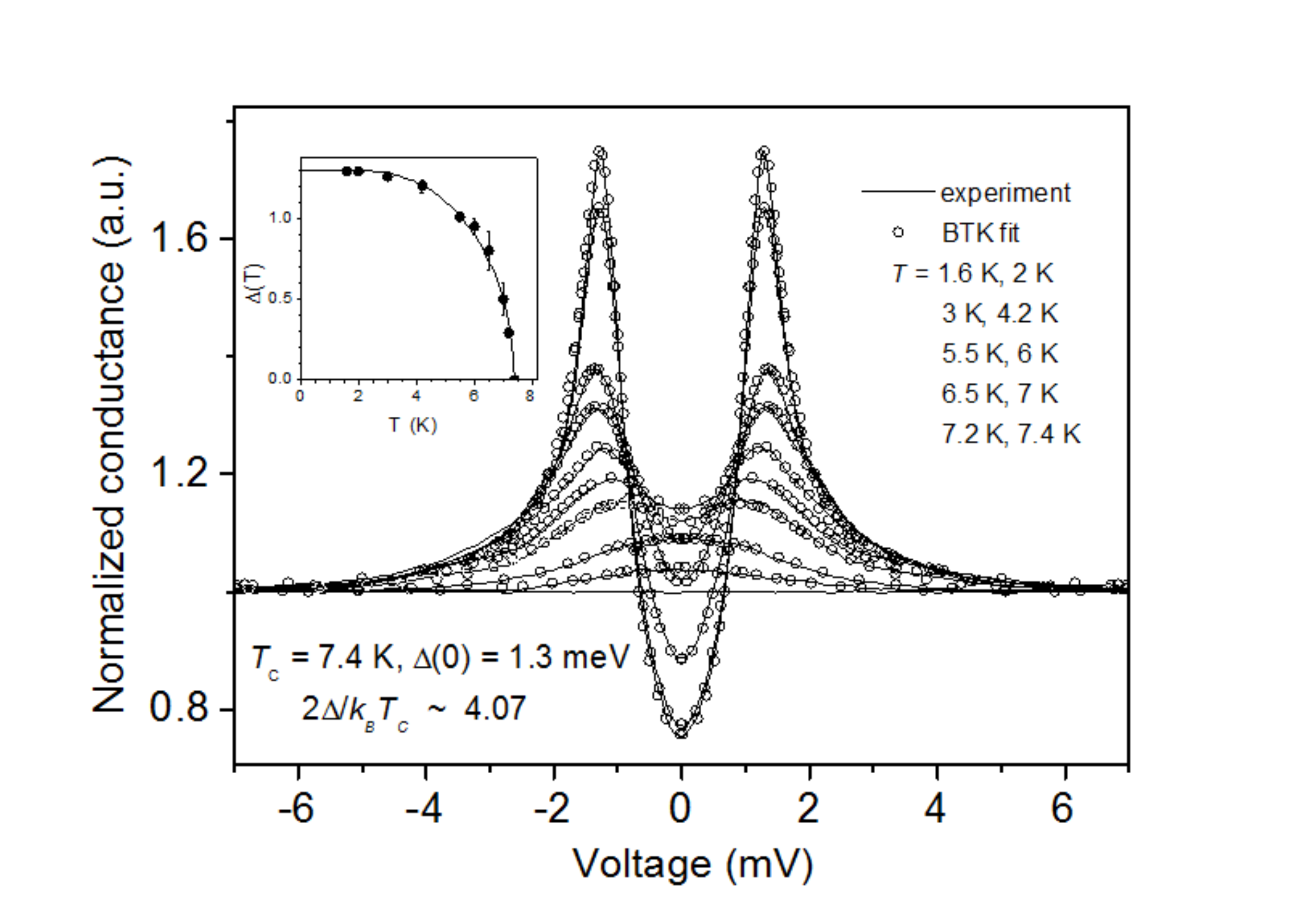}
\caption{Temperature dependence of the high resolution ($\Gamma =0 $) Pt-YB$_6$ point-contact spectrum  (solid lines) measured at indicated temperatures. Open symbols show the BTK 
fits with the fitting parameters $Z=0.64$, and $\Delta(0) = 1.3$ meV. The temperature dependence of the energy gap $\Delta (T)$ is 
shown in the inset by the symbols.  The solid line is the BCS prediction.}
\label{fig:fig2}
\end{figure}

In the following we examined an effect of applied magnetic field on the PC spectra and particularly on the superconducting energy gap. The magnetic field was
 applied perpendicularly to the PC junction area and parallel with a Pt tip. 
The estimated PC diameters were  in the range of hundreds of nanometers while the coherence length in YB$_6$ is about 30 nm. Then, above the lower critical magnetic 
field the junction was in a mixed state with many Abrikosov vortices penetrating the junction.
The vortex cores form a normal-state part $N^A$ of the junction with an area $A$. Their fraction of the whole junction written as 
 $n = N^A/A$ is in a first approximation proportional to the applied field divided to the upper critical one ($H/H_{c2}$). $n$
 increases linearly with magnetic field  and at the upper critical magnetic field where the whole junction is in the normal state $n=1$. 
The normalized point-contact conductance in the mixed state will be the sum $\sigma/\sigma_N(V,H) = n + (1-n)\sigma$, where $n$ represents the
 normal-state channel and $(1-n)\sigma$ is the superconducting-channel contribution. This simple empirical model (successfully applied for the 
study of two-gap superconductivity in MgB$_2$ \cite{bugoslavsky,szabo07}) has been used for the fit of the  normalized-conductance curves measured 
in a presence of magnetic field. Figure 3 shows the evolution of one PC spectrum with magnetic field (lines) and fitting curves (symbols). The fit 
parameters $Z, \Gamma $ and $\Delta $ have been determined from the zero-field PC spectrum. At higher fields  only the values of the field-dependent
 parameters $n$ and $\Delta$ have been varied. The resulting field dependence of the energy gap is plotted by the open circles in the Fig. 5. From relatively small fields it follows  
a square-root dependence, which is in accordance with the Maki's \cite{maki} generalization of the 
Ginzburg-Landau theory for dirty type-II superconductors. 

Similar measurements have been performed at 1.6 K. Thus, $H_{c2}$ values, determined from the PC spectroscopy  as the field where the superconducting features in the PC spectrum vanish, were obtained at these two temperatures.
They are shown in the inset of Fig. 3  by the stars together with the overall   
temperature dependence of the upper critical magnetic field $H_{c2}(T)$  determined from the specific heat measurements performed on the same crystal,
 displayed here as solid squares \cite{specheat}. The consistency 
between $H_{c2}$'s obtained from the bulk-sensitive specific heat measurements and surface sensitive point-contact-spectroscopy measurements points the fact that also the latter reflects the bulk properties and is not affected by any kind of surface degradation. The line over the $H_{c2}(T)$ data points is the classical Werthamer-Helfand-Hohenberg 
model curve \cite{maki}. Our $H_{c2}$ values are in a reasonable agreement with the results of other groups \cite{lortz,kadono}. In particular our
 zero-temperature upper critical field $H_{c2}(0) = 0.28$ T is slightly smaller than 295 mT and 315 mT of Lortz \cite{lortz} and Kadono \cite{kadono}, 
resp., despite our higher $T_c$ than their $T_c=7.1$ K. This indicates that our sample is cleaner with a longer electronic mean free path.

\begin{figure}[t]
\includegraphics[width=8.5 cm]{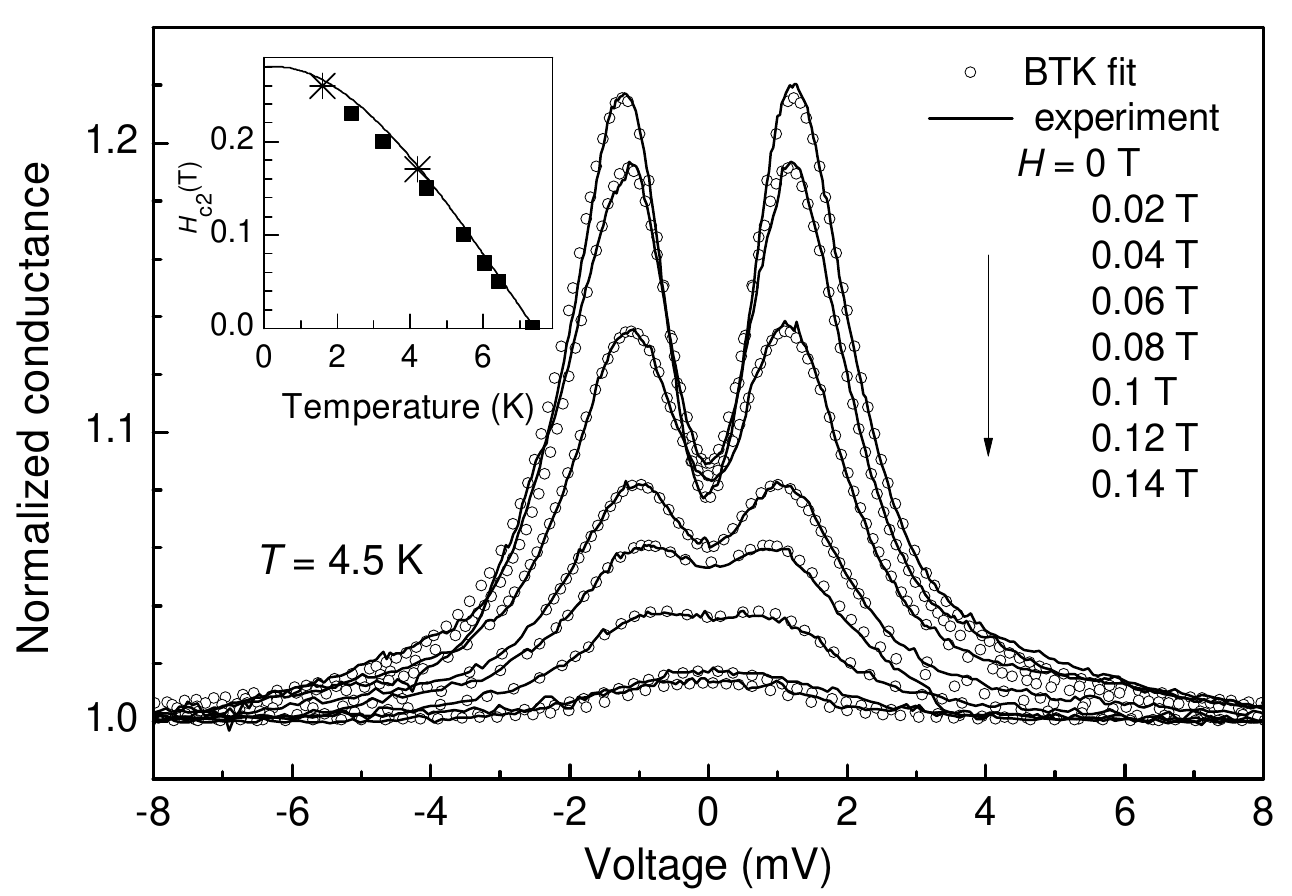}
\caption{Effect of magnetic field on the Pt-YB$_6$  point-contact spectrum (solid lines) measured at $T = 4.5 $ K. Fits by BTK curves in the mixed 
state model (open symbols) yield the parameters $ Z =0.64, \Gamma = 0.23 $ meV and $\Delta$(4.5 K) = 1.15 meV. Inset shows the temperature 
dependence of $H_{c2}$ from our specific heat  (squares) and point-contact measurements (stars). The line is the WHH model.}
\label{fig:fig3}
\end{figure}

\begin{figure}[t]
\includegraphics[width=8.5 cm]{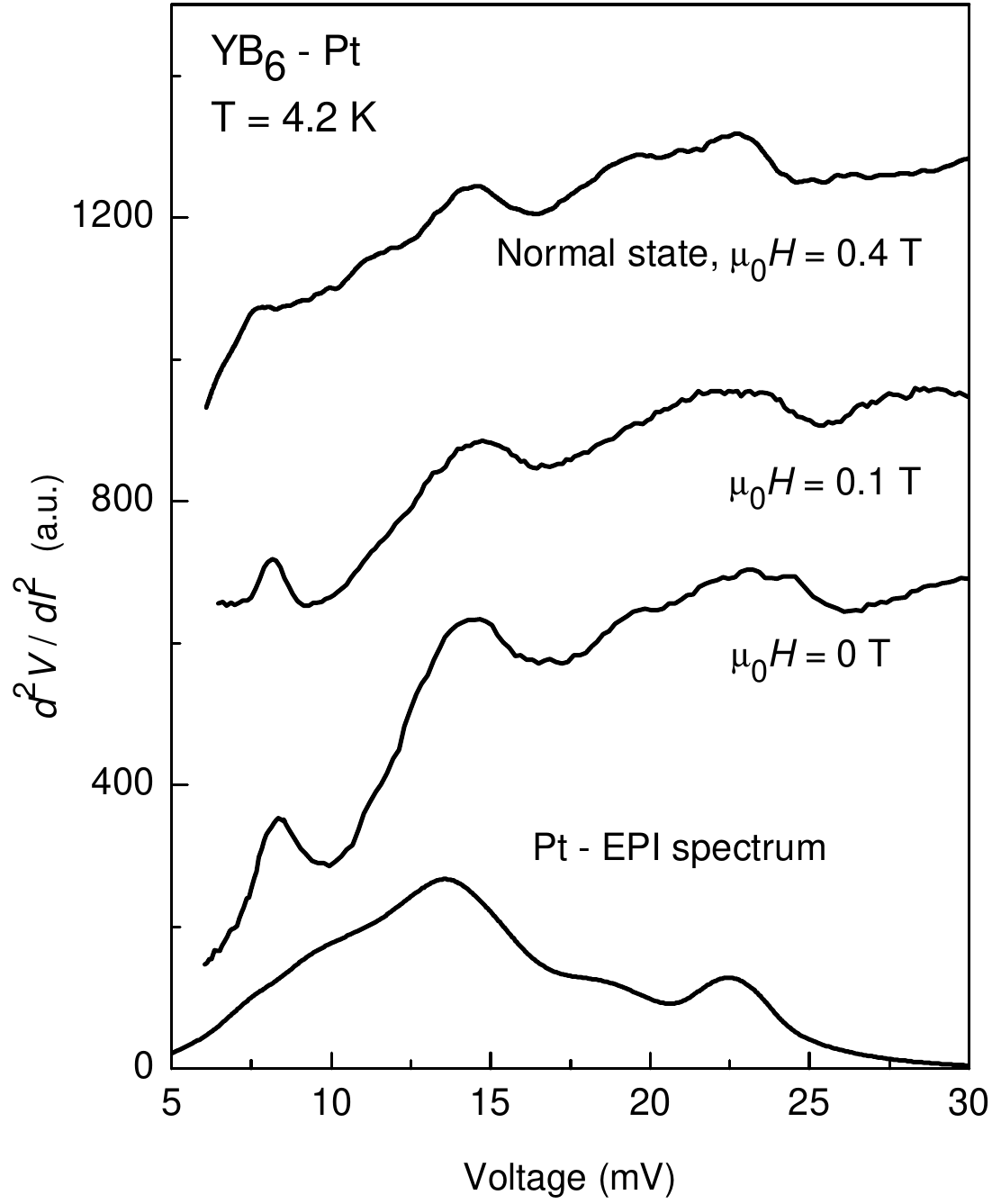}
\caption{Second derivative $d^2V/dI^2(V)$ spectrum measured at $T =$ 4.2 K on a Pt-YB$_6$ point contact in
superconducting (at 0 and 0.1 T) and normal (at 0.4 T) state. The down-most curve plots the EPI function of a Pt-Pt homo-contact  \cite{naidyuk}.}
\label{fig:fig4}
\end{figure}

\begin{figure}[t]
\includegraphics[width=8.5 cm]{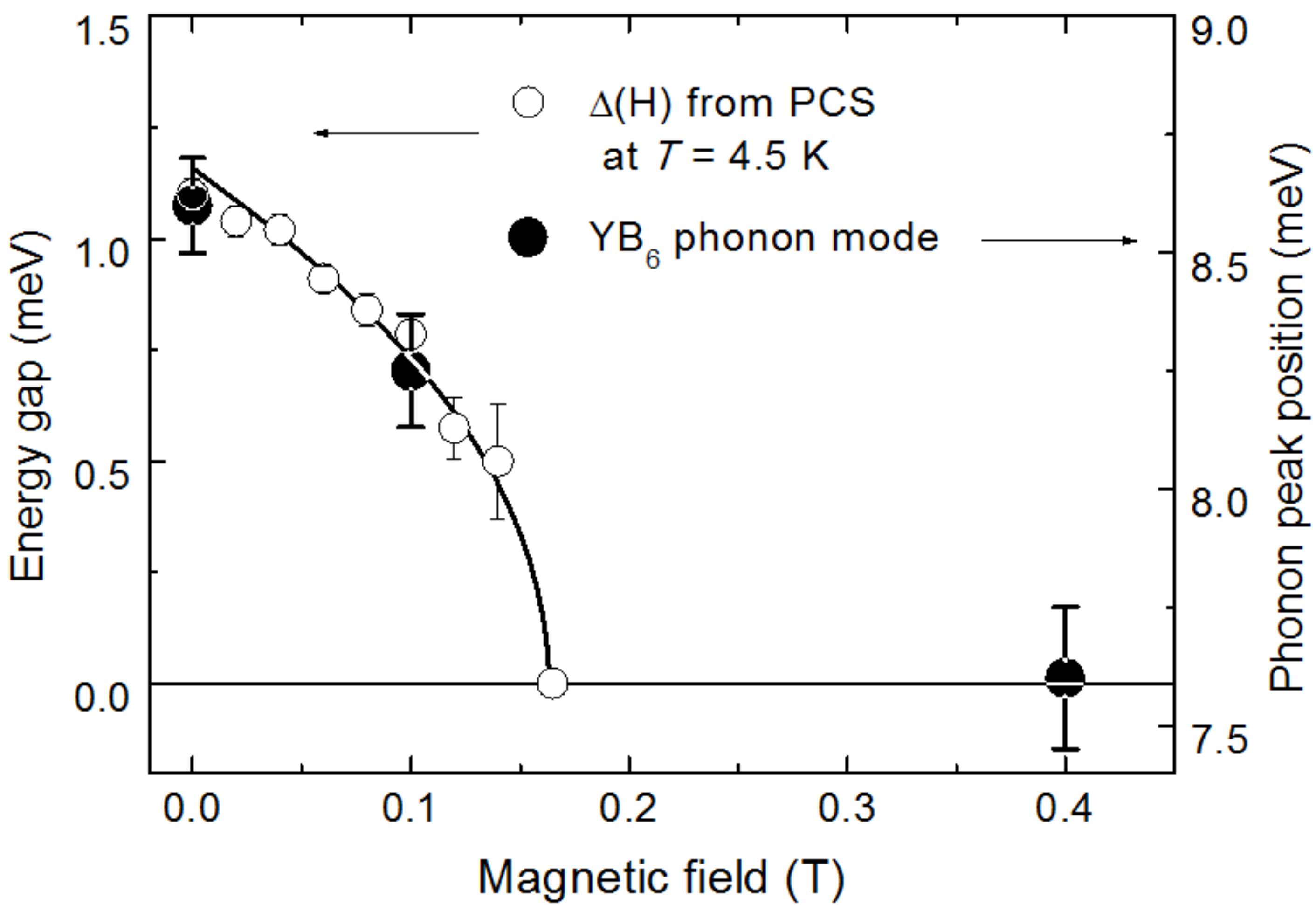}
\caption{Magnetic-field dependence of the superconducting energy gap of YB$_6$ (left ordinate) resulting from the measurements shown in Fig. 3 - 
open circles. The solid line is a square root field dependence. The full circles indicate a position development of the phonon mode of YB$_6$ 
in magnetic field from the spectrum in Fig. 4 (right ordinate applies).}
\label{fig:fig5}
\end{figure}

As has been discussed above, the point-contact spectroscopy is capable to explore the electron-phonon interaction
in  superconductors either due to inelastic  quasi-particle scattering  on phonons, or due to energy dependent superconducting energy gap present in the
elastic  component  of  the  point-contact spectra. In both mechanisms the  second  derivative $d^2V/dI^2(V)$ is
 proportional to the EPI function. The  first mechanism  requires clean ballistic  contacts and
reveals approximately the same spectrum in the superconducting as well as in the normal state. The latter  mechanism  prevails  in the  strongly-coupled
 superconductors, where intensity of the peaks in $ d^2V/dI^2 (V)$ related to the phonons (bosons) mediating the superconducting coupling is significantly 
increased in the superconducting state as compared with the normal state and moreover,   the peak positions are shifted in energies by the value of 
the superconducting energy gap $\Delta$. With a relatively strong superconducting coupling of $2\Delta/k_BT_c = 4 \pm 0.1$ in our YB$_6$ 
samples we expected that strong coupling features could also appear in the elastic component at the characteristic phonon energies. 
By trial and error we looked for the $dV/dI$ spectra showing a 'smooth'
 linearly increasing background, a typical feature of a good quality metallic point contacts with more direct than tunneling conductance. Figure 4
 plots a set of second derivatives $d^2V/dI^2 (V)$ of the point-contact spectra obtained at $T = 4.2$ K in the superconducting as well as the normal 
state of a Pt-YB$_6$ hetero-contact. The curves taken at $H = 0, 0.1$ T show the spectra measured in the superconducting state and mixed state, 
respectively, while the spectrum taken at  $H = 0.4$ T, above the value of the upper critical magnetic field $H_{c2}$(4.2 K) = 0.18 T (see inset of 
 Fig. 3), represents the normal-state behavior. All spectra shown here reveal well defined nonlinearities in the 30 mV window. In the zero-field
 spectrum a sharp peak is visible at the bias (energy) $\omega_1/e$ = 8.6 mV, an intense peak is placed also around $\omega_2/e \approx $ 13 mV 
followed by a hump at about 18 mV and a peak  around $\approx $ 23 mV. The structure is superimposed on the typical point-contact background which 
is related to the nonequilibrium phonon generation near the point-contact orifice \cite{samuely}.

Point-contact spectra measured through a junction between two different metals are proportional to the EPI function of
 both electrodes. For comparison we show in Fig. 4 also the point-contact spectrum of the electron-phonon interaction obtained on the Pt homo-contact, the 
curve is reproduced from Ref. \cite{naidyuk}. In this spectrum the PC background was already subtracted. Two dominant Pt phonon modes are visible
 with peaks at $\sim$ 13 mV and 23 mV and also a hump at about 18 mV can be recognized. Comparing the YB$_6$-Pt zero-field spectrum with the spectrum 
of Pt one can see that the maxima observed above  10 mV are positioned at the characteristic phonon energies of Pt without any energy shift. When the
 superconductivity in our YB$_6$ sample is suppressed in increasing magnetic field,  the intensities and the positions of  these maxima are practically 
unchanged. This is strongly suggestive that in our PC spectra the inelastic scattering of electrons on the Pt phonons dominates at energies above 10 meV. 
On the other hand the low-energy sharp peak observed in the superconducting state at $\omega_1/e \approx 8.6\pm 0.1$ mV changes clearly in applied
  magnetic field. The intensity of this peak is reduced in the increased field and a gradual shift of its position is visible from $\sim $ 8.6 mV
 to $\sim$ 7.6 mV in the normal state, where the peak transforms to a shoulder. 
Such a shift of the energy position between the spectra in the superconducting and normal states indicates that this is the mode mediating superconducting coupling. 
A detailed  comparison of this shit with a progressive  reduction of the superconducting energy gap in magnetic field is shown in Figure 5.  Clearly,  the phonon-peak 
position (solid symbols) follows the suppression of the energy gap $\Delta(H)$ (open symbols) in increasing magnetic field. At higher fields, above $H_{c2}$  the phonon contribution in
 the spectrum due to the energy-dependent superconducting gap is absent and a small non-linearity at about 7.6 mV which is still present as dispalyed in Fig. 4  is due to an 
inelastic quasi-particle scattering on this phonon mode. It is visible together with the phonon modes of the Pt tip.  In contrast to the reports 
in Refs.\cite{kunii,souma} no structure was observed at $\sim 11 $ meV in any of the several junctions revealing detectable non-linearities in the 
 $d^2V/dI^2(V)$  spectra on our Pt-YB$_6$ hetero-contacts. Yet no structures apart of the smoothly increasing background was observed at higher 
voltages above 30 mV.

It is noteworthy that that our measurements have revealed only the strongest peak of the complete EPI spectrum of YB$_6$, while the other branches due to the quasiparticle interactions with the optical 
phonon modes of the boron octahedra are not pronounced. That they have been observed in our PCS studies on the non-superconducting LaB$_6$ \cite{samuely} can be explained by the extremely high quality of the 
LaB$_6$ crystal with the residual resistivity ratio RRR $\cong $ 200 while in the case of YB$_6$ RRR = 4. In the case of our point contact measurements on the LuB$_{12}$ single crystals with RRR = 70 we have observed a dominant EPI peak at about 14 meV coming from Lu vibrations and two small maxima at 24 and 30 meV from optical boron branches \cite{flachb}.

Our finding strongly suggests importance of the low-energy yttrium phonon mode with energy of 7.6 meV in the superconducting coupling of YB$_6$,
 while other phonon modes coming from the boron octahedra vibrations are not significantly coupled to the electronic quasi-particles. This result 
is in a full agreement with the conclusions of the "thermal spectroscopy"  of Lortz \textit{\textit{et al.}} \cite{lortz} based on the heat capacity
 and  resistivity measurements.

\section{Conclusions}

Point-contact spectroscopy measurements have been performed to study the superconducting coupling in YB$_6$.  The values of the superconducting energy gap $\Delta = 1.30 
\pm 0.03$ meV and of the strength of superconductivity $2\Delta / k_BT_c = 4.0 \pm 0.1$ have been determined.  We have shown, that while the maxima of
 $d^2V/dI^2 (V)$ spectra observed on the Pt-YB$_6$ point contacts at energies above 10 mV are related to the inelastic scattering of electrons on the phonons of the Pt tip the 
low-energy mode, observed in the superconducting state at 8.6 mV is shifted to lower energies in the same way as the superconducting energy  gap of YB$_6$ is gradually suppressed 
by the magnetic field and shows up as a small feature at 7.6 meV when
 YB$_6$ is driven to the normal state. This finding is a direct experimental evidence of the phonon-mediated superconductivity 
in YB$_6$ by the low energy yttrium phonon mode near 7.6 meV. Our measurements confirm the results of the deconvolution of the electron-phonon 
interaction from the specific-heat and resistivity measurements by Lortz \textit{et al.} showing that YB$_6$ has been a superconductor
 with an Einstein lattice of Y ions rattling in the spacious cage of boron octahedra.

\begin{acknowledgments}
This work was supported by the following projects: CFNT MVEP - the Centre of Excellence of the Slovak Academy of Sciences, FP7  MNT - ERA.Net II. ESO, 
the EU ERDF (European Union Regional Development Fund) grant No. ITMS26220120005, VEGA No. 2/0135/13 and the APVV-0036-11 grant of the Slovak R\&D Agency. 
The liquid nitrogen for the experiment has been sponsored by the U.S. Steel Ko\v sice, s.r.o.
\end{acknowledgments}

\end{document}